\documentclass[twocolumn,showpacs,preprintnumbers,amsmath,amssymb]{revtex4}
\usepackage{dcolumn}
\usepackage{bm}

\begin {document}

\title {On strong superadditivity for a class of quantum channels}

\author {Grigori G. Amosov}

\email {gramos@mail.ru}

\affiliation {Department of Higher Mathematics\\
Moscow Institute of Physics and Technology\\ Dolgoprudny
141700\\RUSSIA}

\date{\today}

\begin {abstract}
Given a quantum channel $\Phi $ in a Hilbert space $H$ put $\hat
H_{\Phi}(\rho)=\min \limits _{\rho _{av}=\rho }\Sigma
_{j=1}^{k}\pi _{j}S(\Phi (\rho _{j}))$, where $\rho _{av}=\Sigma
_{j=1}^{k}\pi _{j}\rho _{j}$, the minimum is taken over all
probability distributions $\pi =\{\pi _{j}\}$ and states $\rho
_{j}$ in $H$, $S(\rho)=-Tr\rho\log\rho$ is the von Neumann
entropy of a state $\rho$. The strong superadditivity conjecture
states that $\hat H_{\Phi \otimes \Psi}(\rho)\ge \hat
H_{\Phi}(Tr_{K}(\rho))+\hat H_{\Psi}(Tr_{H}(\rho))$ for two
channels $\Phi $ and $\Psi $ in Hilbert spaces $H$ and $K$,
respectively. We have proved the strong superadditivity
conjecture for the quantum depolarizing channel in prime
dimensions. The estimation of the quantity $\hat H_{\Phi\otimes
\Psi}(\rho)$ for the special class of Weyl channels $\Phi $ of the
form $\Phi=\Xi \circ \Phi_{dep}$, where $\Phi _{dep}$ is the
quantum depolarizing channel and $\Xi $ is the phase damping is
given.

\end {abstract}


\maketitle

\section {Introduction}

A linear trace-preserving map $\Phi $ on the set of states
(positive unit-trace operators) $\mathfrak{S}(H)$ in a Hilbert
space $H$ is said to be a quantum channel if $\Phi ^{*}$ is
completely positive (\cite {Hol}). The channel $\Phi $ is called
bistochastic if $\Phi (\frac {1}{d}I_{H})=\frac {1}{d}I_{H}$.
Here and in the following we denote by $d$ and $I_{H}$ the
dimension of $H,\ dimH=d<+\infty ,$ and the identity operator in
$H$, respectively.

Given a quantum channel $\Phi $ in a Hilbert space $H$ put (\cite
{Sh})
\begin {equation}\label {quant}
\hat H_{\Phi}(\rho)=\min \limits _{\rho _{av}=\rho}\sum \limits
_{j=1}^{k}\pi _{j}S(\Phi (\rho _{j})),
\end {equation}
where $\rho _{av}=\sum \limits _{j=1}^{k}\pi _{j}\rho _{j}$ and
the minimum is taken over all probability distributions $\pi
=\{\pi _{j}\}$ and states $\rho _{j}\in \mathfrak{S}(H)$. Here
and in the following $S(\rho)=-Tr(\rho\log \rho)$ is the von
Neumann entropy of a state $\rho $. The strong superadditivity
conjecture states that
\begin {equation}\label {strong}
\hat H_{\Phi \otimes \Psi}(\rho)\ge \hat
H_{\Phi}(Tr_{K}(\rho))+\hat H_{\Psi}(Tr_{H}(\rho)),
\end {equation}
$\rho \in \mathfrak{S}(H\otimes K)$ for two channels $\Phi $ and
$\Psi $ in Hilbert spaces $H$ and $K$, respectively.

The infimum of the output entropy of a quantum channel $\Phi $ is
defined by the formula
\begin {equation}\label {addit}
\chi (\Phi)=\inf \limits _{\rho\in \mathfrak{S}(H)}S(\Phi (\rho)).
\end {equation}
The additivity conjecture for the quantity $\chi (\Phi)$ states
(\cite {Hol2})
$$
\chi (\Phi\otimes \Psi)=\chi (\Phi)+\chi (\Psi)
$$
for an arbitrary quantum channel $\Psi $. It was shown in (\cite
{Sh}) that if the strong superadditivity conjecture holds, then
the additivity conjecture for the quantity $\chi $ holds too.
Nevertheless the conjecture (\ref {strong}) is stronger than
(\ref {addit}).

In the present paper we shall prove the strong superadditivity
conjecture for the quantum depolarizing channel in prime
dimensions of $H$. We also give some estimation from below for
the quantity $\hat H_{\Phi \otimes \Psi}(\rho)$ for the certain
class of Weyl channels $\Phi $.

\section {The Weyl channels}

Fix the basis $|f_{j}>\equiv |j>,\ 0\le j\le d-1,$ of the Hilbert
space $H$. We shall consider a special subclass of the
bistochastic Weyl channels (\cite {Amo, Amo1, Ruskai, Fukuda,
Cerf}) defined by the formula (\cite {Amo1})
\begin {equation}\label {Weyl}
\Phi (\rho)=(1-(d-1)(r+dp))\rho +r\sum \limits
_{m=1}^{d-1}W_{m,0}\rho W_{m,0}^{*}
\end {equation}
$$
+p\sum \limits _{m=0}^{d-1}\sum \limits _{n=1}^{d-1}W_{m,n}\rho
W_{m,n}^{*},
$$
$\rho\in \mathfrak{S}(H)$, where $r,p\ge 0,\ (d-1)(r+dp)=1$ and
the Weyl operators $W_{m,n}$ are determined as follows
$$
W_{m,n}=\sum \limits _{k=0}^{d-1}e^{\frac {2\pi i}{d}kn}|k+m\ mod\
d
><k|,
$$
$0\le m,n\le d-1$.

 Consider the maximum commutative group ${\mathcal U}_{d}$
consisting of unitary operators
$$
U=\sum \limits _{j=0}^{d-1}e^{i\phi _{j}}|e_{j}><e_{j}|,
$$
where the orthonormal basis $(e_{j})$ is defined by the formula
$$
|e_{j}>=\frac {1}{\sqrt d}\sum \limits _{k=0}^{d-1}e^{\frac {2\pi
i}{d}jk}|k>,\ 0\le j\le d-1,
$$
$\phi _{j}\in {\mathbb R},\ 0\le j\le d-1$. Notice that
$$
<f_{k}|e_{j}>=\frac {1}{\sqrt d}e^{\frac {2\pi i}{d}jk},\ 0\le
j,k\le d-1,
$$
It implies that
\begin {equation}\label {MUB}
|<f_{k}|e_{j}>|=\frac {1}{\sqrt d}
\end {equation}
The bases $(f_{j})$ and $(e_{j})$ satisfying the property (\ref
{MUB}) are said to be mutually unbiased (\cite {Ivan}). It is
straightforward to check that
\begin {equation}\label {shift}
W_{0,n}|e_{j}><e_{j}|W_{0,n}^{*}=|e_{j+n\ mod\ d}><e_{j+n\ mod\
d}|,
\end {equation}
$0\le j,n\le d-1$.

It was shown in \cite {Amo1} that the Weyl channels (\ref {Weyl})
are covariant with respect to the group ${\mathcal U}_{d}$ such
that
$$
\Phi (UxU^{*})=U\Phi (x)U^{*},\ x\in \sigma (H),\ U\in {\mathcal
U}_{d}.
$$

{\bf Example 1.} Put $r=p=\frac {q}{d^{2}},\ 0\le q\le 1$, then it
can be shown (\cite {Amo, Amo1, Ruskai}) that (\ref {Weyl}) is the
quantum depolarizing channel,
\begin {equation}\label {dep}
\Phi _{dep}(\rho)=(1-q)\rho +\frac {q}{d}I_{H},\ \rho\in
\mathfrak{S}(H),
\end {equation}
$$
\chi (\Phi _{dep})=-(1-\frac {d-1}{d}q)\log (1-\frac
{d-1}{d}q)-(d-1)\frac {q}{d}\log \frac {q}{d}.
$$
$\Box $

{\bf Example 2.} Put $r=\frac {1}{d}(1-\frac {d-1}{d}q),\ p=\frac
{q}{d^{2}},\ 0\le q\le \frac {d}{d-1}$, then (\ref {Weyl}) is
q-c-channel (\cite {Hol2}). Indeed, under the conditions given
above the channel $\Phi \equiv \Phi _{qc}$ can be represented as
follows
$$
\Phi _{qc}(\rho)=(1-\frac {d-1}{d}q)E(\rho)+\frac {q}{d}\sum
\limits _{n=1}^{d-1}W_{0,n}E(\rho)W_{0,n},
$$
where
$$
E(\rho)=\frac {1}{d}\sum \limits _{m=0}^{d-1}W_{m,0}\rho
W_{m,0}^{*},
$$
$\rho\in \mathfrak{S}(H)$ is a conditional expectation on the
algebra generated by the projections $|e_{j}><e_{j}|,\ 0\le j\le
d-1$. Taking into account (\ref {shift}) we get
\begin {equation}\label {qc}
\Phi _{qc}(\rho)=\sum \limits
_{j=0}^{d-1}Tr(|e_{j}><e_{j}|\rho)\rho_{j},\ \rho\in
\mathfrak{S}(H),
\end {equation}
where
$$
\rho_{j}=(1-\frac {d-1}{d}q)|e_{j}><e_{j}|+
$$
$$
\frac {q}{d}\sum \limits _{k=1}^{d-1}|e_{j+k\ mod\ d}><e_{j+k\
mod\ d}|,
$$
$0\le j\le d-1$,
$$
\chi (\Phi _{qc})=-(1-\frac {d-1}{d}q)\log (1-\frac
{d-1}{d}q)-(d-1)\frac {q}{d}\log \frac {q}{d}.
$$

$\Box $

{\bf Proposition 1.}{\it Suppose that the channel $\Phi $ has the
form (\ref {Weyl}) and $p\le r\le \frac {1}{d}(1-d(d-1)p)$. Then,
it can be represented as
$$
\Phi =\lambda \Phi _{dep}+(1-\lambda)\Phi _{qc},
$$
$0\le \lambda \le 1$, where $\Phi _{dep}$ and $\Phi _{qc}$ are
defined by the formulae (\ref {dep}) and (\ref {qc}),
respectively.}

Proof.

It follows from the condition $p\le r\le \frac {1}{d}(1-d(d-1)p)$
that there exists a number $\lambda ,\ 0\le \lambda \le 1,$ such
that $r=\lambda p+(1-\lambda)\frac {1}{d}(1-d(d-1)p)$.

$\Box $

Suppose that the powers $U^{k}$ of a unitary operator $U$ in a
Hilbert space $H$ form a cyclic group of the order $d$. Fix the
probability distribution $\pi =\{\pi _{k},\ 0\le k\le d-1\}$,
then the bistochastic quantum channel $\Xi $ defined by the
formula
$$
\Xi (\rho)=\sum \limits _{k=0}^{d-1}\pi _{k}U^{k}\rho U^{*k},\
\rho \in \mathfrak{S}(H),
$$
is said to be {\it a phase damping}.

{\bf Proposition 2.}{\it Suppose that the channel $\Phi $ has the
form (\ref {Weyl}) and $p\le r\le \frac {1}{d}(1-d(d-1)p)$, then
\begin {equation}\label {comp}
\Phi (\rho)=\Xi \circ \Phi
_{dep}(\rho),\ \rho \in \mathfrak{S}(H),
\end {equation}
where $\Phi _{dep}$ is the quantum depolarizing channel (\ref
{dep}) and $\Xi $ is the phase damping defined by the formula
$$
\Xi (\rho)=\frac {1+(d-1)\lambda }{d}\rho +\frac {1-\lambda}{d}
\sum \limits _{m=1}^{d-1}W_{m,0}\rho W_{m,0}^{*},\ \rho \in
\mathfrak{S}(H),
$$
$0\le \lambda \le 1$. }

{\bf Remark.} {\it The additivity conjecture for channels of the
form (\ref {comp}) was proved in \cite {Amo}.}

Proof.

It is sufficiently to pick up the number $\lambda $ defined in
Proposition 1.

$\Box $

\section {The estimation of the output entropy}

Our approach is based upon the estimate of the output entropy
proved in \cite {Amo1}. Here we shall formulate the corresponding
theorem without a proof for the convenience.

{\bf Theorem 2 (\cite {Amo1}).}{\it Let $\Phi
(\rho)=(1-p)\rho+\frac {p}{d}I_{H},\ \rho\in \mathfrak{S} (H),\
0\le p\le \frac {d^{2}}{d^{2}-1},$ be the quantum depolarizing
channel in the Hilbert space $H$ of the prime dimension $d$.
Then, there exist $d$ orthonormal bases $\{e_{j}^{s},\ 0\le s,j\le
d-1\}$ in $H$ such that
\begin {equation}\label {XJ}
S((\Phi \otimes Id)(\rho))\ge -(1-\frac {d-1}{d}p)\log (1-\frac
{d-1}{d}p)-
\end {equation}
$$
\frac {d-1}{d}p\log \frac {p}{d}+ \frac {1}{d^{2}}\sum \limits
_{j=0}^{d-1}\sum \limits _{s=0}^{d-1}S(\rho_{j}^{s}),
$$
where $\rho\in \mathfrak{S} (H\otimes K),\
\rho_{j}^{s}=dTr_{H}((|e_{j}^{s}><e_{j}^{s}|\otimes I_{K})\rho)\in
\mathfrak {S} (K),\ 0\le j,s\le d-1$. }

In the present paper our goal is to prove the following theorem.

{\bf Theorem.}{\it Let $\Phi $ be the Weyl channel (\ref {Weyl})
in the Hilbert space of the prime dimension $d$ satisfying the
property $p\le r\le \frac {1}{d}(1-d(d-1)p)$. Then, for an
arbitrary quantum channel $\Psi $ in a Hilbert space $K$ the
inequality
$$
\hat H_{\Phi \otimes \Psi}(\rho)\ge -(1-\frac {d-1}{d}p)\log
(1-\frac {d-1}{d}p)-
$$
$$
\frac {d-1}{d}p\log \frac {p}{d}+\hat H_{\Psi}(Tr_{H}(\rho)),\
\rho \in \mathfrak{S}(H\otimes K),
$$
holds.}

{\bf Remark.}{\it Due to the covariance property of $\Phi _{dep}$
we get
$$
\hat H_{\Phi _{dep}}(\rho)=-(1-\frac {d-1}{d}p)\log (1-\frac
{d-1}{d}p)-
$$
$$
\frac {d-1}{d}p\log \frac {p}{d}=const.
$$
Hence, the theorem implies that
$$
\hat H_{\Phi _{dep}\otimes \Psi}(\rho)\ge \hat H_{\Phi _{dep}
}(Tr_{K}(\rho))+\hat H_{\Psi}(Tr_{H}(\rho)),
$$
$\rho \in \mathfrak{S}(H\otimes K).$ }

Proof.

At first, let us prove the theorem only for the quantum
depolarizing channel $\Phi _{dep}$. Put $\tilde \rho =(Id\otimes
\Psi)(\rho)$.

It follows from Theorem 2 of \cite {Amo1} that
$$
S((\Phi _{dep}\otimes Id)(\tilde \rho))\ge -(1-\frac
{d-1}{d}p)\log (1-\frac {d-1}{d}p)-
$$
$$
\frac {d-1}{d}p\log \frac {p}{d}+ \frac {1}{d^{2}}\sum \limits
_{j=0}^{d-1}\sum \limits _{s=0}^{d-1}S(\rho_{j}^{s}),
$$
where $\rho\in \mathfrak {S} (H\otimes K),\
\rho_{j}^{s}=dTr_{H}((|e_{j}^{s}><e_{j}^{s}|\otimes I_{K})\tilde
\rho)\in \mathfrak {S} (K),\ 0\le j,s\le d-1$.

Notice that
\begin {equation}\label {Sup1}
\frac {1}{d^{2}}\sum \limits _{j=0}^{d-1}\sum \limits
_{s=0}^{d-1}S(\rho_{j}^{s})=Tr_{H}(\tilde \rho)=\Psi
(Tr_{H}(\rho)).
\end {equation}
It follows from equality (\ref {Sup1}) that
$$
\frac {1}{d^{2}}\sum \limits _{j=0}^{d-1}\sum \limits
_{s=0}^{d-1}S(\rho_{j}^{s})\ge \hat H_{\Psi}(Tr_{H}(\rho))
$$
and we have proved the strong superadditivity conjecture for the
quantum depolarizing channel.

The Weyl channel $\Phi $ satisfying the conditions of Theorem can
be represented as a composition
$$
\Phi =\Xi \circ \Phi _{dep}
$$
in virtue of Proposition 2. It implies that
$$
\hat H_{\Phi \otimes \Psi}(\rho)\ge \hat H_{\Phi _{dep}\otimes
\Psi}(\rho)
$$
due to the non-decreasing property of the von Neumann entropy.
Thus, the result follows from the strong superadditivity property
of the quantum depolarizing channel we have proved above.

$\Box $

\section {Conclusion}

We have shown that our method introduced in \cite {Amo, Amo1,
Amo2} allows to prove the strong superadditivity conjecture for
the quantum depolarizing channel. This method based upon the
decreasing property of the relative entropy doesn't use the
properties of $l_{p}$-norms of quantum channels. Thus, we suppose
that the approach is fruitful for the future investigations in
quantum information theory.


\begin {thebibliography}{99}

\bibitem {Amo} Amosov G.G. Remark on the additivity conjecture for
the depolarizing quantum channel. Probl. Inf. Transm. 42 (2006)
3-11. e-print quant-ph/0408004.

\bibitem {Amo1} Amosov G.G. On the Weyl  channels being covariant
with respect to the maximum commutative group of unitaries.
e-print quant-ph/0605177 v.3.


\bibitem {Amo2} Amosov G.G. On the additivity conjecture for the
Weyl channels being covariant with respect to the maximum
commutative group of unitaries. e-print quant-ph/0606040 v.3.


\bibitem {AHW} Amosov G.G., Holevo A.S., Werner R.F. On some additivity problems in
quantum information theory.  Probl. Inf. Transm. 2000. V. 36. N 4.
P. 24-34; e-print quant-ph/0003002.

\bibitem {Ruskai} Datta N, Ruskai M.B. Maximal output purity and capacity for asymmetric unital qudit channels
J. Physics A: Mathematical and General 38 (2005) 9785-9802.
e-print quant-ph/0505048.

\bibitem {Fukuda} Fukuda M., Holevo A.S. On Weyl-covariant
channels. e-print quant-ph/0510148.


\bibitem {Hol} Holevo A.S. On the mathematical theory of quantum communication
channels. Probl. Inf. Transm. 8 (1972) 62 - 71.

\bibitem {Hol1} Holevo A.S. Some estimates for the amount of information
transmittable by a quantum communications channel. (Russian)
Probl. Inf. Transm. 9 (1973) 3 - 11.

\bibitem {Hol2} Holevo A.S. Quantum coding theorems. Russ.
Math. Surveys 53 (1998) 1295-1331; e-print quant-ph/9808023.


\bibitem {Sh} Holevo A.S., Shirokov M.E. On Shor's channel
extension and constrained channels. Commun. Math. Phys. 249
(2004) 417-436.

\bibitem {Ivan} Ivanovich I.D. Geometrical description of quantum state
determination. J. Physics A 14 (1981) 3241-3245.


\bibitem {Cerf} Karpov E., Daems D., Cerf N.J. Entanglement
enhanced classical capacity of quantum communication channels
with correlated noise in arbitrary dimensions. e-print
quant-ph/0603286.

\end {thebibliography}
\end {document}